# Electrohydrodynamic Wind Generation in Planar DBDs: Role of Electrode Symmetry and Geometry

Farshad Sohbatzadeh[1, 2, *], Saeed Ranjbar Malekshah[1], Hamed Soltani Ahmadi[1, 2], S. Mirzanejhad[1,2], Ramin Mehrabifard[3], Samira Mavvadati[4], Zdenko Machala[3]

[1]Department of Atomic and Molecular Physics, Faculty of Science, University of Mazandaran, Babolsar, 47416-95447, Mazandaran, Iran

[2]Plasma Technology Research Core, Faculty of Science, University of Mazandaran, Babolsar, Iran

[3]Division of Environmental Physics, Faculty of Mathematics, Physics and Informatics, Comenius University, Mlynska dolina, 842 48 Bratislava, Slovakia

[4]Electronic Department, Faculty of Engineering and Technology, University of Mazandaran, Babolsar, Iran

[*]Corresponding Author: f.sohbat@umz.ac.ir

**Abstract**

This study experimentally and numerically investigates the electrohydrodynamic (EHD) interaction induced by a surface dielectric barrier discharge (SDBD) actuator at atmospheric pressure. The EHD effect, driven by non-thermal plasma in a dielectric barrier discharge (DBD), generates ionic wind, which is characterized here for a symmetric annular-type DBD actuator. A symmetric annular-type DBD actuator, consisting of concentric ring–disk electrodes that generate a predominantly vertical ionic wind rather than a tangential jet. Despite extensive studies on linear and tangential SDBD actuators, the influence of annular electrode geometry on vertically induced ionic wind and associated ozone generation remains insufficiently explored. We analyze the induced wind velocity perpendicular to the electrode plane, focusing on the influence of geometric parameters—electrode diameter (D) and thickness (δ)—on performance. Experimental results reveal a maximum wind velocity of 3.42 m/s ± 1% for an optimized configuration (D = 32 mm, δ = 0.06 mm), corroborated by numerical simulations. The simulations further elucidate the velocity profile, volumetric force, electron temperature, and gas pressure distribution within the plasma region. Complementary diagnostics, including $O_3$ concentration measurements and Schlieren imaging, demonstrate that larger electrode diameters (e.g., 32 mm and 22 mm) enhance vertical flow height but concurrently increase ozone production. These findings provide actionable insights for designing DBD-based systems in applications such as plasma flow control, air purification, and biomedical plasma technologies.

## 1. Introduction

Dielectric barrier discharge (DBD) was first pioneered by Siemens in 1857 for ozone generation [1]. Today, DBD plasmas—operating at non-thermal, atmospheric-pressure conditions—are widely employed across scientific and industrial fields, including cancer therapy [2–6], bacterial inactivation [7,8], surface modification [9–11], and air purification [12–14]. DBDs exhibit three characteristic modes: filamentary (random micro-discharges), glow-like (homogeneous plasma), and patterned (surface-interactive structures) [15].

Surface DBD (SDBDs have received considerable interest in recent years due to their prospective uses in biological, environmental, and agricultural fields. In most of these applications, the plasma does not directly touch the substrate getting treatment, and the transfer of reactive species from the plasma to the substrate is often supposed to be controlled by diffusion [16,17]. Despite the uniform construction of all SDBDs, characterized by two electrodes divided by a dielectric barrier, many electrode designs, including square [18], hexagonal[19] , circular [20], and spiral, have been suggested in recent years [21]. This study focuses on SDBD actuators, which generate ionic wind perpendicular to the electrode plane. Early SDBD configurations often employed asymmetric linear electrode arrangements to generate tangential wall jets. This concept was later extended to plasma synthetic jet actuators (PSJAs)[22]. SDBD plasma actuators have been used in a variety of sectors outside of aviation due to their unique characteristics [23]. They have shown efficacy in wind turbine blades, flame holders, combustion [24–26], enhancement of lift in vehicles[24] [27], and regulation of tip clearance flow in turbines [28,29]. These varied applications illustrate the versatility and potential of plasma actuators across several sectors. The use of active control flow methods may enhance performance in controlled systems. The voltage waveform critically influences actuator’s performance. Kotsonis and Ghaemi [27][30] demonstrated that square waveforms yield stronger ionic winds than sinusoidal signals. Similarly, annular PSJAs outperform linear designs in energy conversion efficiency [22,31–35]. Beyond aerodynamics, DBD actuators enhance combustion processes [36–39] and turbulent drag reduction. For instance, Zheng et al. [40] showed that annular DBD actuators (A-DBD-PAs) stabilize wall-normal jets, reducing shear stress by 30% in turbulent boundary layers. Despite these advances, the impact of electrode geometry-particularly diameter and thickness—on annular SDBD performance remains underexplored.

. However, despite these advances, the impact of annular SDBD electrode geometry particularly electrode diameter (D) and thickness (δ) on vertical ionic-wind performance remains underexplored. In this work, we combine experiments and numerical modelling to quantify the wall-normal (perpendicular) ionic wind produced by a symmetric annular SDBD actuator as a function of D and δ, and to identify a geometry that maximizes the induced flow. The simulations provide insight into the spatial distributions of the electrohydrodynamic (EHD) volumetric force, electron temperature, and gas pressure in the plasma region. Finally, Schlieren imaging and ozone measurements are used to evaluate practical trade. Since ozone is a regulated respiratory irritant with strict occupational and ambient-air thresholds, increased ozone production with larger electrodes is a relevant practical constraint for operation in occupied or enclosed environments. These findings provide actionable design guidelines for high-efficiency DBD actuators, with applications spanning aerodynamic flow control, air purification, and biomedical plasma devices.

## 2. Simulation and experimental setup

### 2.1. The structure of the SDBD

The surface dielectric barrier discharge (SDBD) actuator in this study features a circular symmetric geometry, comprising two concentric aluminum electrodes separated by a glass dielectric layer. The upper electrode is a ring-shaped aluminum foil with varying diameters (D) and thicknesses (δ), while the lower electrode is a circular aluminum disk. Key geometric parameters include upper electrode width fixed at 3 mm for all configurations, and glass dielectric layer at 4 mm thick. The electrode gap maintained at 5 mm. The lower electrode (grounded disk) was fabricated from aluminum and, for consistency across configurations, its material type and thickness were chosen to match those of the upper electrode (i.e., the same foil/tape or sheet and the same thickness δ for each case). The dielectric thickness, inter-electrode gap, and electrode width were fixed to limit the parameter space and isolate the influence of D and δ. These baseline dimensions were selected as a literature-consistent and fabrication-practical compromise to ensure stable discharge operation at kV excitation (not independently optimized here), noting that dielectric thickness and electrode spacing are known to affect charge storage, discharge morphology, and EHD forcing [41]. The aluminum electrodes, with high electrical conductivity (≈$3.8\times10^7$ S/m), facilitate strong electric fields and efficient charge injection. The glass dielectric layer (relative permittivity ≈5–7, dielectric strength ≈13–20 kV/mm) enables surface charge accumulation and discharge stabilization. Electrode surface roughness (on the micrometer scale) and the materials' thermal properties further influence discharge uniformity and thermal management during sustained operation. These parameters are critical for the geometric design and overall performance of the plasma actuator. Figure 1 illustrates the schematic and physical prototype of the SDBD actuator, while Table 1 summarizes the geometric variations tested for the upper electrode. We used electrode diameters of 14, 22, 32, 42, and 54 millimeters, and electrode thicknesses of 0.06, 0.10, 0.50, 1.00, 2.00, and 4.00 millimeters.

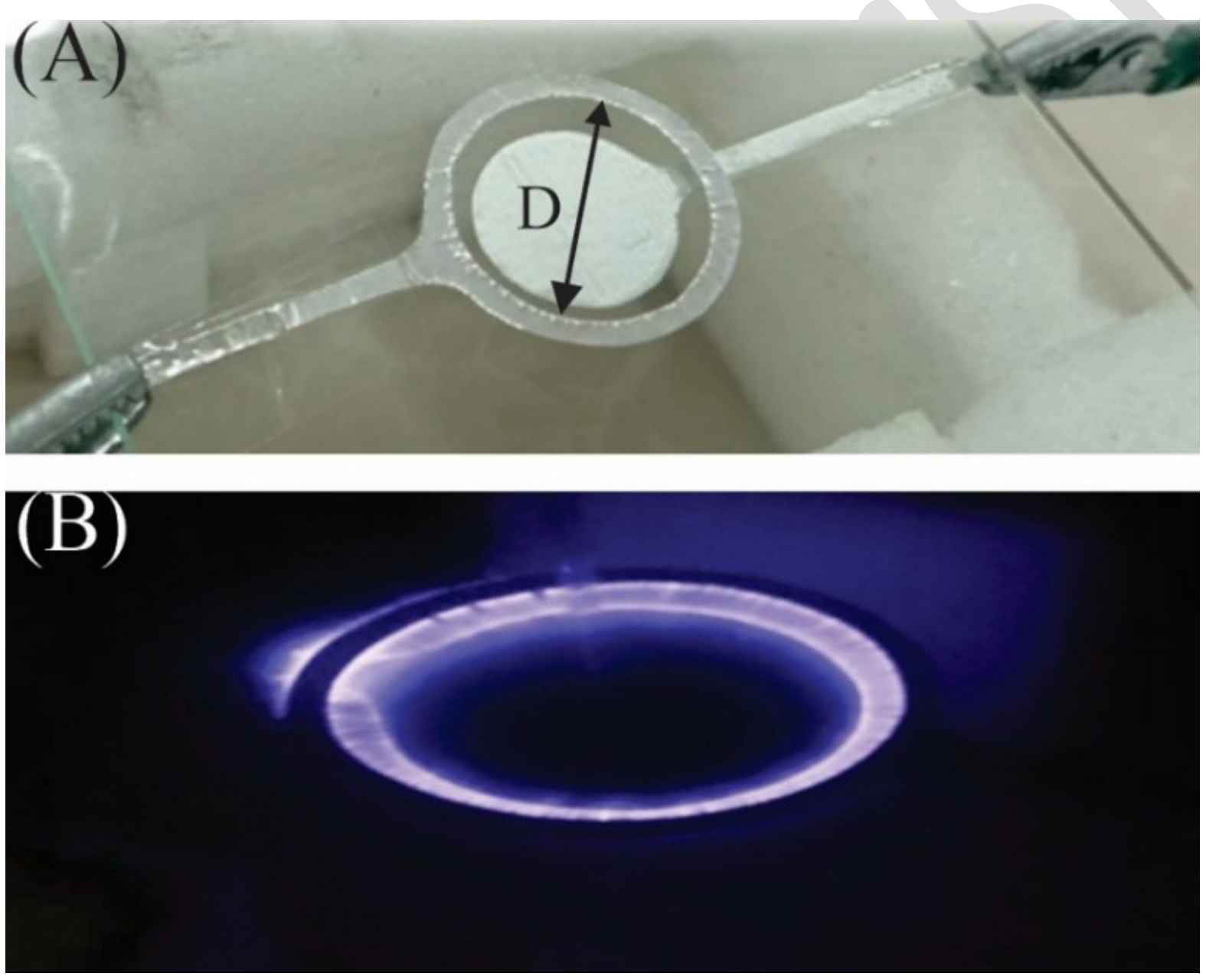

Figure 1. (A) The real image of the ring electrodes of the SDBD. (B) The image of ignited SDBD at a frequency of 18.9 kHz with a peak-to-peak voltage of 16 kV.

Table 1. Geometrical characteristics and material of the manufactured annular electrodes

| Electrode number | Diameter (mm) | Thickness (mm) | Electrode material |
|---|---|---|---|
| 1 | 14 ± 0.5 | 0.06 ± 0.005 | Aluminum Adhesive Tape |
| 2 | 22 ± 0.5 | 0.06 ± 0.005 | |
| 3 | 32 ± 0.5 | 0.06 ± 0.005 | |
| 4 | 32 ± 0.5 | 0.50 ± 0.01 | Aluminum Sheet |
| 5 | 32 ± 0.5 | 1.00 ± 0.01 | |

| 6 | 32 ± 0.5 | 2.00 ± 0.01 | |
|---|---|---|---|
| 7 | 32 ± 0.5 | 4.00 ± 0.01 | |
| 8 | 42 ± 0.5 | 0.06 ± 0.005 | Aluminum Adhesive Tape |
| 9 | 54 ± 0.5 | 0.10 ±0.005 | Aluminum Sheet |
| 10 | 54 ± 0.5 | 1.00 ± 0.01 | |
| 11 | 54 ± 0.5 | 2.00 ± 0.01 | |
| 12 | 54 ± 0.5 | 4.00 ± 0.01 | |

The 54-mm upper electrode was fabricated from a 0.1-mm aluminum foil rather than the ~0.06-mm foil used for smaller diameters, solely to ensure mechanical rigidity at the larger diameter. Although this change was not intended to modify electrical behavior, a thicker foil can, in principle, influence the system in several ways: (i) increased electrode mass loading can slightly reduce vibrational sensitivity, (ii) the greater edge thickness may alter the local curvature of the electric field near the electrode boundary, and (iii) the modified edge shape may weakly influence the onset or spatial distribution of microdischarges. In the present work, we did not observe any qualitative differences in discharge appearance for the 54-mm electrode relative to other diameters, and the effects of electrode thickness are separately evaluated in the δ-sweep experiments. Therefore, results attributed to diameter changes (D-dependence) are not conflated with unintended thickness effects.

**2.2 Simulation and boundary condition**

To analyze the charged particle distribution and electrohydrodynamic (EHD) force effects on fluid motion in the SDBD system, we conducted a two-dimensional numerical simulation of ring electrodes based on prior methodologies [16,42]. The simulation was implemented in COMSOL Multiphysics (v6.2) with a computational domain of 800 × 400 mm (Figure 2). The model comprises three aluminum electrodes: two positioned on the upper dielectric surface and a grounded electrode on the lower surface. The 32 mm spacing between the upper electrodes matches the optimal annular electrode diameter from experimental tests. The dielectric material (glass) replicates the experimental setup. A non-uniform mesh was employed to resolve strong plasma gradients, with refined triangular elements near the discharge region and coarser elements elsewhere. Boundary conditions and surface properties are detailed in Table 2. This approach ensures numerical convergence while resolving the steep near-electrode gradients of electric field and charged-species densities that determine the EHD body force (ρE) driving the neutral-gas flow.

Limitations of the 2D simplification. The present COMSOL model is two-dimensional and therefore cannot represent all inherently three-dimensional aspects of an annular SDBD actuator and its induced flow. Two-dimensional SDBD/DBD fluid models are widely used to capture the dominant electrostatic/plasma processes and the resulting EHD body force distribution in the electrode cross-section; however, they inherently neglect azimuthal non-uniformities and 3D flow structures. In annular and axisymmetric DBD/PSJA configurations, the induced jet can involve toroidal vortices, entrainment, and circumferential variations of discharge intensity that are not captured in a planar 2D model. Consequently, the 2D model is expected to provide reliable qualitative trends and spatial localization of EHD forcing, but may differ quantitatively in peak velocity, jet spreading, and local current fluctuations compared to the full 3D system. In this work, we therefore use the 2D simulation primarily to interpret the measured trends and distributions (EHD force, electron temperature, pressure), and we benchmark key outputs (current waveform and the location of the velocity maximum) against experiments.

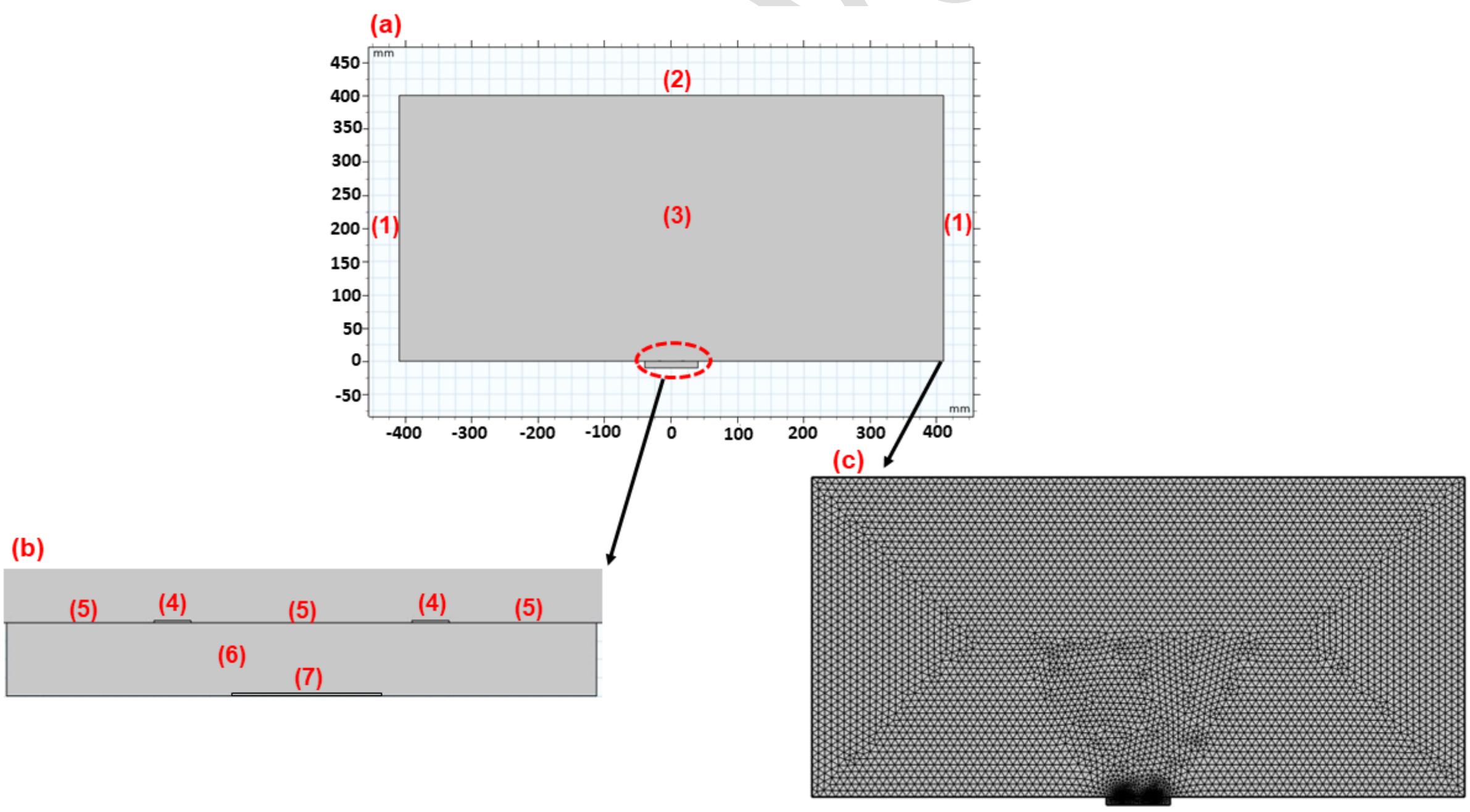

Figure 2. (a) Overall computational region, (b) close-up view of the plasma formation region, (c) meshing structure

As specified in Table 2, periodic boundary conditions (PBC) were employed to approximate an infinite system. The surface charge accumulation boundary condition accounts for charged particle deposition at dielectric interfaces, modifying the initial electric field. Wall boundaries incorporate surface reactions and combine open reactions. The upper electrodes were driven by a sinusoidal voltage source $V_1 =$ $V \times \sin(\omega_0 \times t)$, where V = 16 kV and $\omega_0 = 2\pi f$ with f = 20 kHz, while the lower electrode remained grounded ($V_0 = 0$). A settling time of $t = 1\times10^{-4}$ s (two periods) ensured stable plasma formation. Charge conservation (Region 6) maintained constant total electric charge within the dielectric. Boundaries 1 and 2 represented the gas inlet (velocity = 0.001 m/s) and outlet, respectively. For computational efficiency, the working gas was simplified to pure nitrogen, representing the dominant (80%) component of air.

Table 2. Boundary conditions and computational region of different parts of the simulation system

| **Boundary or Domain conditions** | **Boundary or Domain number** |
|---|---|
| Periodic Boundary Condition (PBC) | 1 and 2 |
| Surface Charge Accumulation | 5 |
| Wall Conditions | 4 and 5 |
| Ground ($V=V_0$) | 7 |
| Metal Contact (V=V1) | 4 |
| Charge Conservation (Dielectric) | 6 |
| Gas inlet | 1 |
| Gas outlet | 2 |
| Volume Force | 3 |
| Surface reactions | 4 and 5 |
| Electron cross-section reaction and reaction between heavy species | 3 |

Periodic Boundary Conditions (PBC) applied to boundaries 1 and 2 simulate an infinitely repeating domain in the horizontal direction, effectively reducing computational cost while modeling a spatially extended system. Dirichlet conditions are applied to boundaries with fixed potentials: the upper electrodes (Boundary 4, V=V1) and the lower grounded electrode (Boundary 7, V=V0). These define the applied voltage driving the discharge. Neumann and specialized conditions model surface interactions. The Surface Charge Accumulation condition (Boundary 5) on the dielectric surface allows for charge deposition, modifying the local electric field—a key feature of dielectric barrier discharges. Wall Conditions (Boundaries 4 and 5) incorporate surface recombination and secondary electron emission processes. Charge Conservation within the dielectric domain (Region 6) ensures the conservation of total charge. These conditions collectively govern the electric field distribution, charge transport, and plasma-fluid coupling

The numerical simulations employ COMSOL's fluid (drift-diffusion) plasma model, which treats the plasma as a quasi-neutral continuum rather than tracking individual particles. This approach utilizes the finite element method (FEM) to discretize the computational domain and numerically solve the governing equations [43–45]. The simulation incorporates all relevant reaction mechanisms, including electron impact reactions, surface reactions, and nitrogen-based reactions, as detailed in previous studies [43,44,46–48]. The initial simulation parameters are summarized in Table 3.

Table 3. Initial simulation parameters.

| Initial parameter | Parameter value |
|---|---|
| Pressure | 760 [Torr] |
| Temperature | 300 [K] |
| Initial mean electron energy | 5 [V] |
| Initial electron density | $10^{12}$ [1/m$^3$] |
| Electron mobility | Calculate from the electron collision cross section |

The electrohydrodynamic (EHD) volume force in domain 3 is given by:

$$F_{Coulomb} = \rho E \quad (1)$$

Where, $F_{Coulomb}$ is the Coulomb force density (N/m³), ρ represents the net charge density (C/m³), and E denotes the electric field intensity (V/m). The Coulomb force term (ρE) dominates the EHD interaction, driving fluid motion through momentum transfer from charged particles to neutral molecules. This force arises from the coupling between the plasma's space charge (ρ) and the applied electric field (E), generating the characteristic ionic wind in SDBD actuators.

### 2.3. Electrical measurements

The discharge electrical parameters were measured using two Tektronix P6015A high-voltage probes (1000× attenuation, 75 MHz bandwidth), and two current monitors (Pearson model 4100, 1 V/A sensitivity) to capture discharge current. A Gwinstek GDS-3354 digital oscilloscope (4-channel, 300 MHz bandwidth, 5 GS/s sampling rate) for simultaneous signal acquisition. The plasma power was determined through time-resolved electrical measurements. The instantaneous power P(t) was calculated as:

$$P(t) = V(t).I(t) \quad (2)$$

Where V(t) is the instantaneous voltage, and I(t) is the instantaneous current. The average discharge power was obtained by integrating the instantaneous power over one complete AC cycle

$$P_{avg} = \frac{1}{T}\int_0^T P(t)dt = \frac{1}{T}\int_0^T V(t).I(t)\,dt \quad (3)$$

Where T is the period of the applied voltage waveform.

### 2.4. Velocity measurement and Pitot tube

A Pitot tube was employed to measure the time-resolved fluid flow rate induced by the plasma discharge (Figure 3). The induced electric wind velocity was quantified at varying positions along the X and Y axes using a calibrated micromanometer (CEM, DT-8920. PONPE, Inc), with an accuracy of ±0.01 m/s. The device was calibrated to a static pressure standard at regular intervals before the experiments began. To mitigate sparking during measurements, a glass Pitot tube (with an inner diameter of 0.5 mm and an outer diameter of 1 mm) replaced the standard stainless-steel tube. The length to diameter ratio of the pipe was observed to ensure the complete development of the flow. For each point (a combination of radial distance and elevation), data were collected during 5 repetitions in a 60-second interval, and their mean and standard deviation were calculated. Also, all experiments were performed at constant temperature and controlled relative humidity to reduce the effect of air density changes on Pitot readings.

Static pressure was measured at a wall-mounted tap located 18 cm upstream of the actuator and outside the region influenced by the ionic wind, ensuring an undisturbed reference pressure. This value was used as the baseline (ambient) static pressure for all Pitot-differential measurements.

Prior to the plasma-actuator measurements, the Pitot tube and differential manometer were checked under low-speed, steady flow conditions to ensure accurate performance in the low-dynamic-pressure regime characteristic of EHD-induced velocities. First, a zero-offset check was performed with both ports open to ambient to verify that the baseline pressure reading returned to zero. Second, the differential-pressure response was compared against a reference manometer for small pressure differences representative of the expected velocity range. Finally, the pressure–velocity conversion was validated in a laminar duct flow of known cross-section and volumetric flow rate, confirming that the measured differential pressure produced the correct mean velocity. These steps ensure that the Pitot/manometer system provides reliable time-averaged velocity measurements for the low-speed, oscillation-prone EHD flows examined in this work.

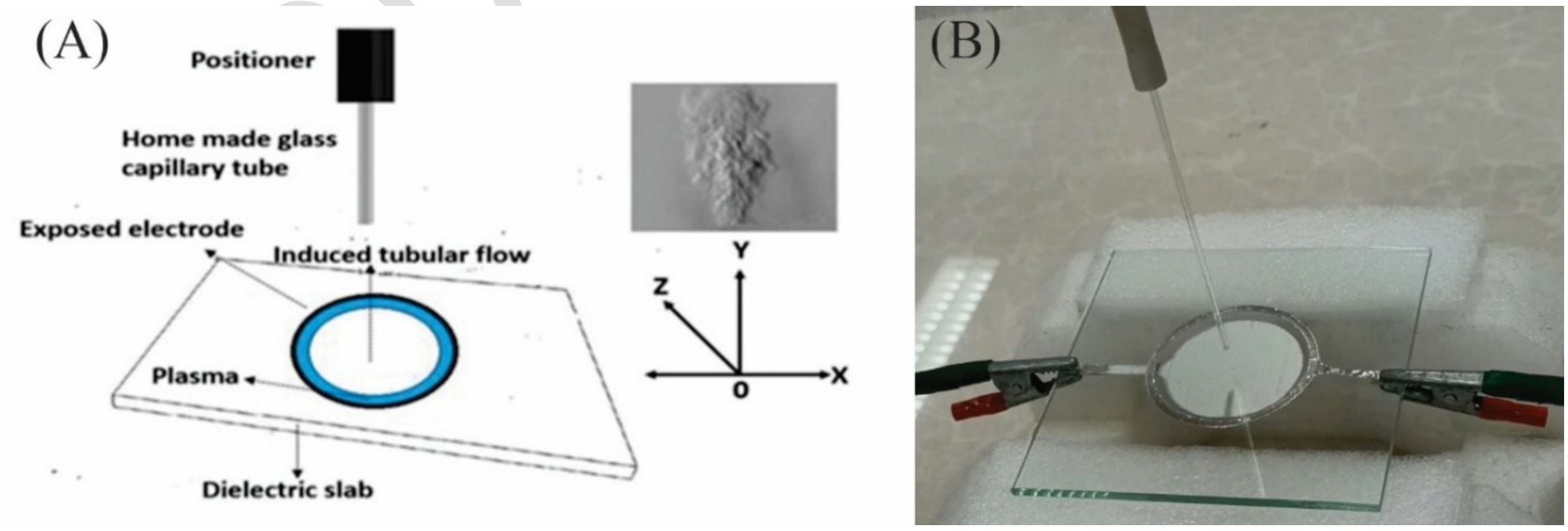

Figure 3. (A) Schematic and, (B) real picture of measuring electric wind velocity perpendicular to the surface electrode plane.

**2.5. Measurement of ozone produced by the annular actuators**

Given the ozone-generating capability of dielectric barrier discharge (DBD) systems, ozone concentration was quantified under varying experimental conditions. An ozone meter sensor (Ozone sensor 0-10ppm v1.0, DFROBOT, Inc.) was positioned 18.0 cm ± 0.05 cm above the DBD electrode (Figure 4), and measurements were conducted for electrodes with diameters of 14, 22, 32, 42, and 54 mm, maintaining a fixed thickness of 0.06 mm. All tests were performed under consistent discharge conditions of Peak-to-peak voltage at 16 kV, and frequency at 18.9 kHz. The accuracy of the ozone meter is 1 ppb. For each electrode, after the plasma was lit, ozone concentrations were recorded for 130 second at 10-second intervals until it reached a stable state. Intervals of 2 to 3 hours were maintained between tests to allow ozone concentrations in the environment to revert to their initial state. The reported number is the average stable data. In order to ensure that ionic wind flow did not affect the sensor reading, a comparative test was performed by placing the sensor in a small sampling chamber with a calm flow and no significant difference was observed.
The velocity field generated by plasma-induced ionic wind can exhibit unsteady fluctuations at the excitation frequency and strong spatial gradients close to the exposed electrode. In the present study, the Pitot tube was connected to a differential manometer whose inherent response time is on the order of tenths of a second. As a result, the measurement acts as a low-pass temporal filter and provides a time-averaged (quasi-steady) velocity rather than a phase-resolved waveform synchronized to the AC voltage. At each measurement location, readings were recorded only after the signal stabilized, and the displayed mean value was used for analysis. Because of the finite probe diameter and the steep gradients within the first few millimeters of the wall, the near-electrode peak velocity may be partially spatially smoothed; therefore, the reported profiles should be interpreted as cycle-averaged, locally averaged velocities.

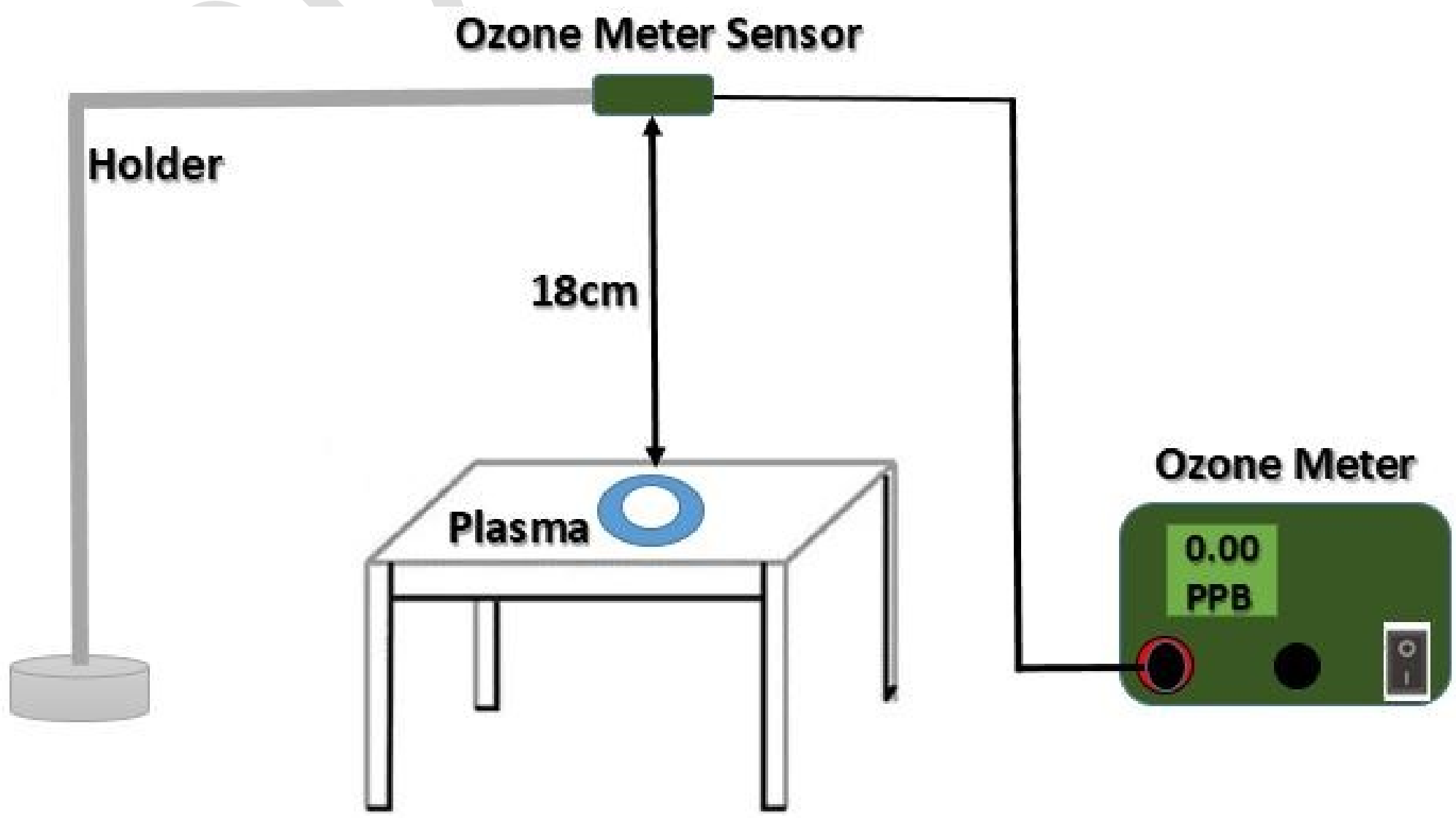

Figure 4. Schematic of the ozone meter and how its sensor is placed above the actuator.

Ozone concentration was sampled at a fixed height of 18 cm above the actuator surface. This location was chosen because it lies outside the immediate near-electrode region, where ozone and flow fields exhibit very steep gradients, yet remains close enough to capture geometry-dependent differences in ozone transport. Since the purpose of the present measurement was comparative (evaluating changes in ozone output among actuator configurations under identical operating conditions), a single representative height was used consistently for all tests. We acknowledge that ozone concentration in vertically rising EHD-driven flows may vary with height; therefore, the reported values correspond to this specific sampling location.

### 2.6. Schlieren imaging

The Schlieren imaging system employed a single-mirror configuration as illustrated in Figure 5. An LED light source was collimated and spatially filtered through a pinhole aperture to generate a point light source. This incoherent light was directed towards a concave mirror and reflected back through the measurement plane. At the focal plane, a sharp-edged blade served as the Schlieren stop, selectively obstructing portions of the light beam to produce diffraction patterns. The modified light field was captured using a CCD camera (Model RA-ASC-0267) with a frame rate of 8.1 Hz, and an adjustable exposure time from microseconds to milliseconds. Image acquisition and processing were performed using the manufacturer's proprietary software. This configuration allowed for visualization and analysis of the diffraction phenomena generated by the blade edge.

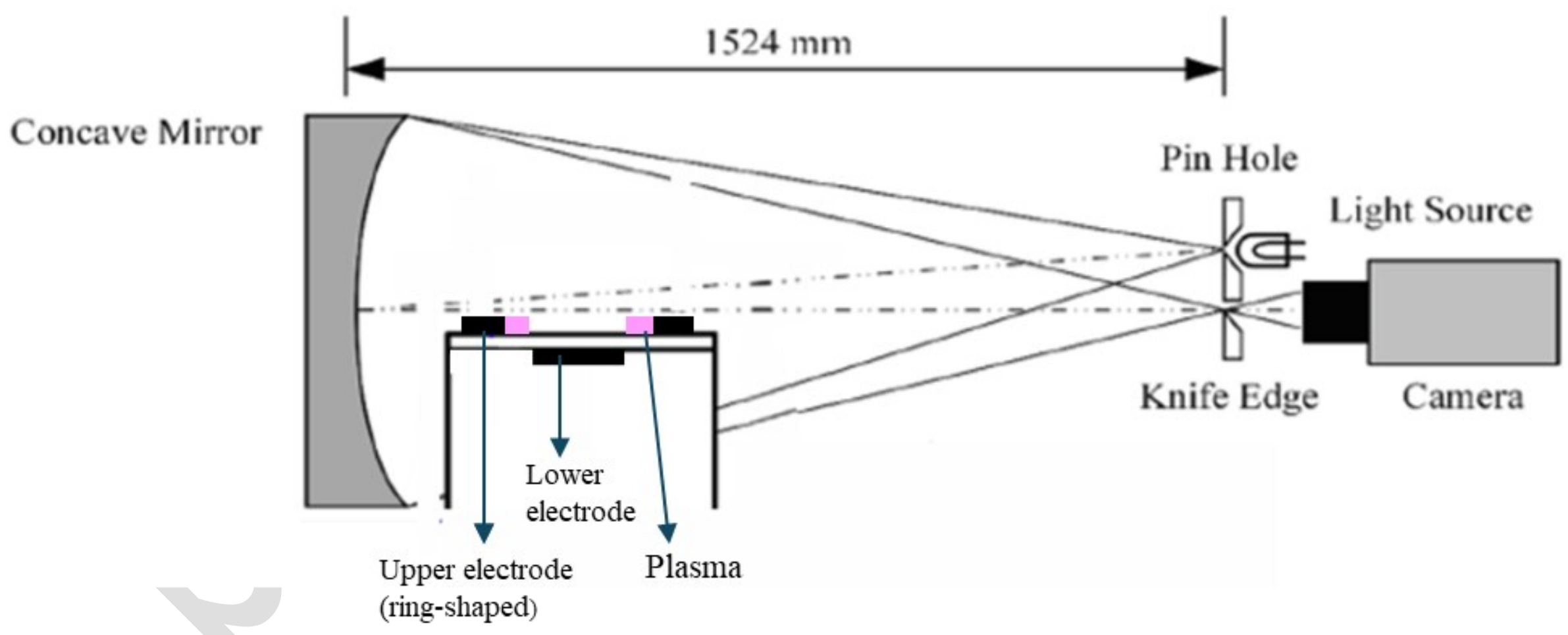

Figure 5. Cross-section schematic of Schlieren imaging.

## 3. Experimental results and simulations

### 3.1- Ionic wind velocity

The induced ion wind velocity profiles were measured using a calibrated Pitot tube system. Velocity measurements were first conducted across electrodes of varying diameters to determine the optimal configuration for vertical flow generation. Subsequently, electrodes with the optimal diameter were fabricated at different thicknesses to further characterize their effect on vertical wind velocity. This two-stage optimization process enabled identification of the electrode geometry producing maximum vertical induced flow.

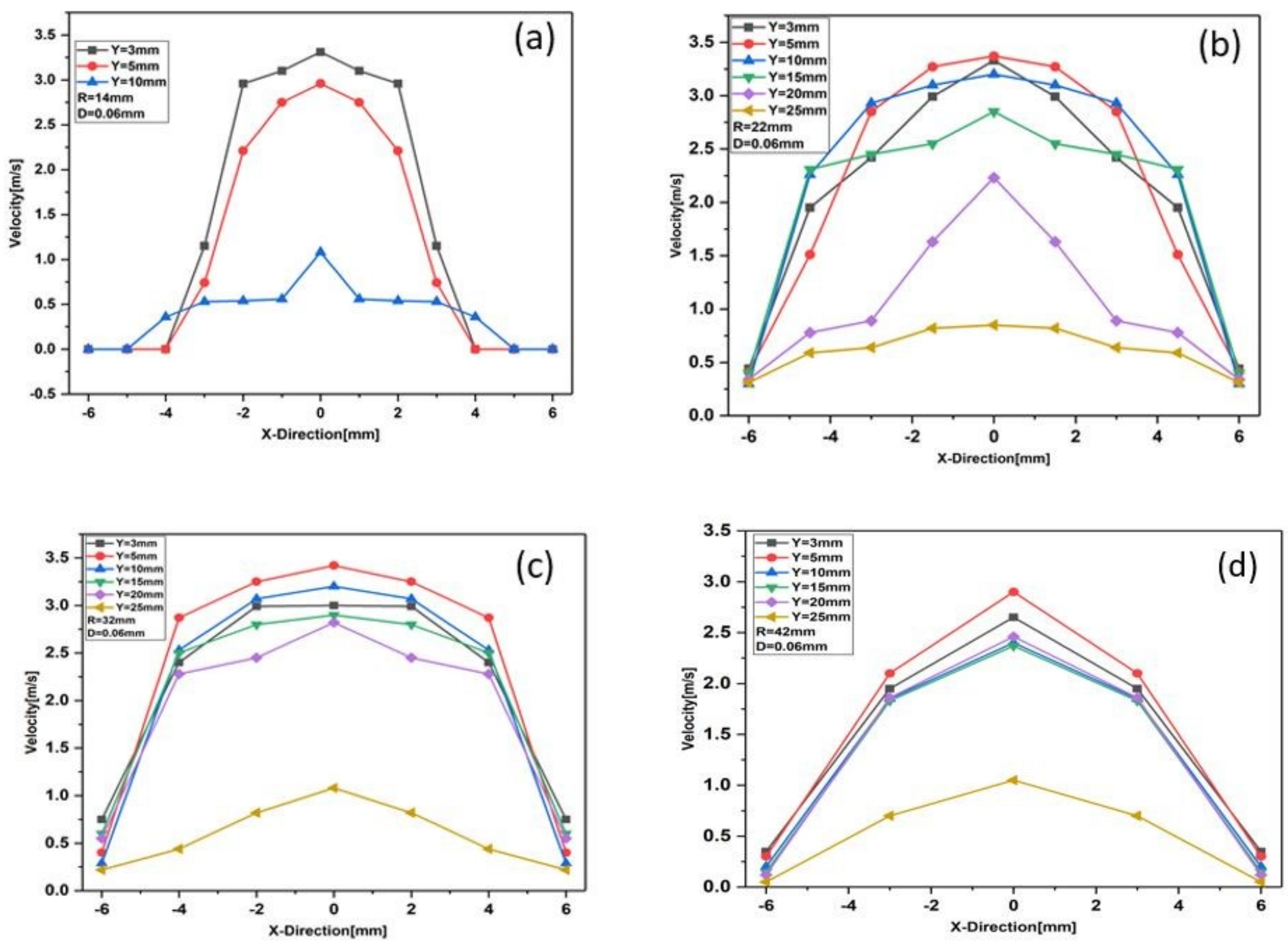

Figure 6. Changes in ion wind velocity profile in electrode thickness of 0.06 mm and electrode diameter (a) 14 mm (b) 22 mm (c) 32 mm (d) 42 mm. The velocity after averaging and with a relative standard deviation of 1% were provided.

At this stage, all electrodes had a constant thickness of 0.06 mm, while their diameters varied. Figure 6 presents the average velocities measured at different radial distances and heights for each electrode. For the 14 mm diameter electrode (D = 14 mm), the maximum wind velocity 3.31 m/s ± 1%) occurred at the center (height: 3.00 mm), decreasing radially until reaching zero (Figure 6-a). The 22 mm diameter electrode (D = 22 mm) exhibited a broader velocity profile, with a peak velocity of 3.37 m/s ± 1% at the center and a height of 5.00 mm (Figure 6-b). In the 32 mm diameter electrode (D = 32 mm), the optimal velocity profile was observed at 5.00 mm height (Figure 6-c). Similarly, the 42 mm electrode achieved its maximum velocity (2.90 m/s ± 1%) at 5.00 mm height (Figure 6-d). Due to fabrication constraints, the 54 mm electrode was constructed with a 0.10 mm -thick aluminum foil instead of 0.06 mm. Its velocity profile is shown separately in Figure 7.

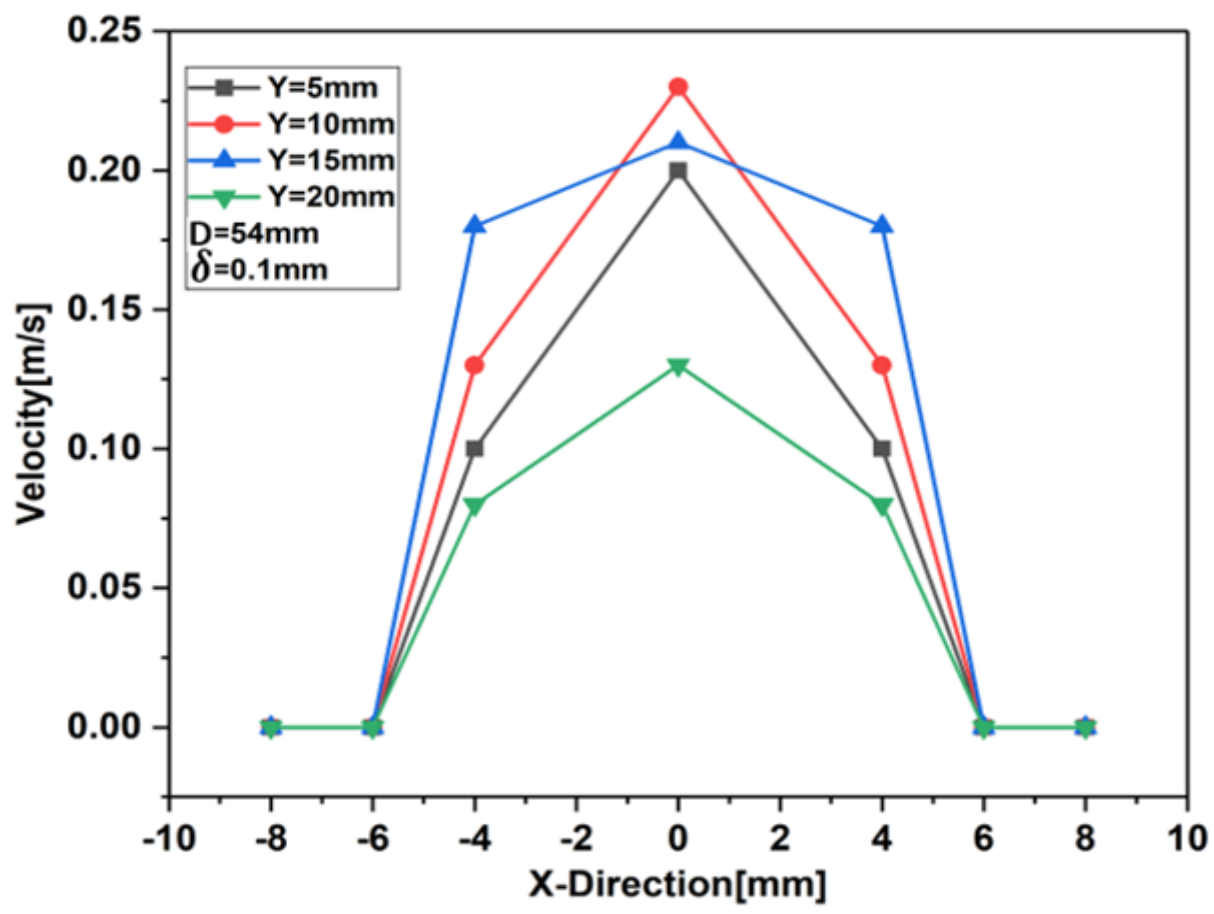

Figure 7. Changes in ion wind velocity profile in a ring electrode with a thickness of 0.10 mm and a diameter of 54 mm. The velocity after averaging and with a relative standard deviation of 1% were provided

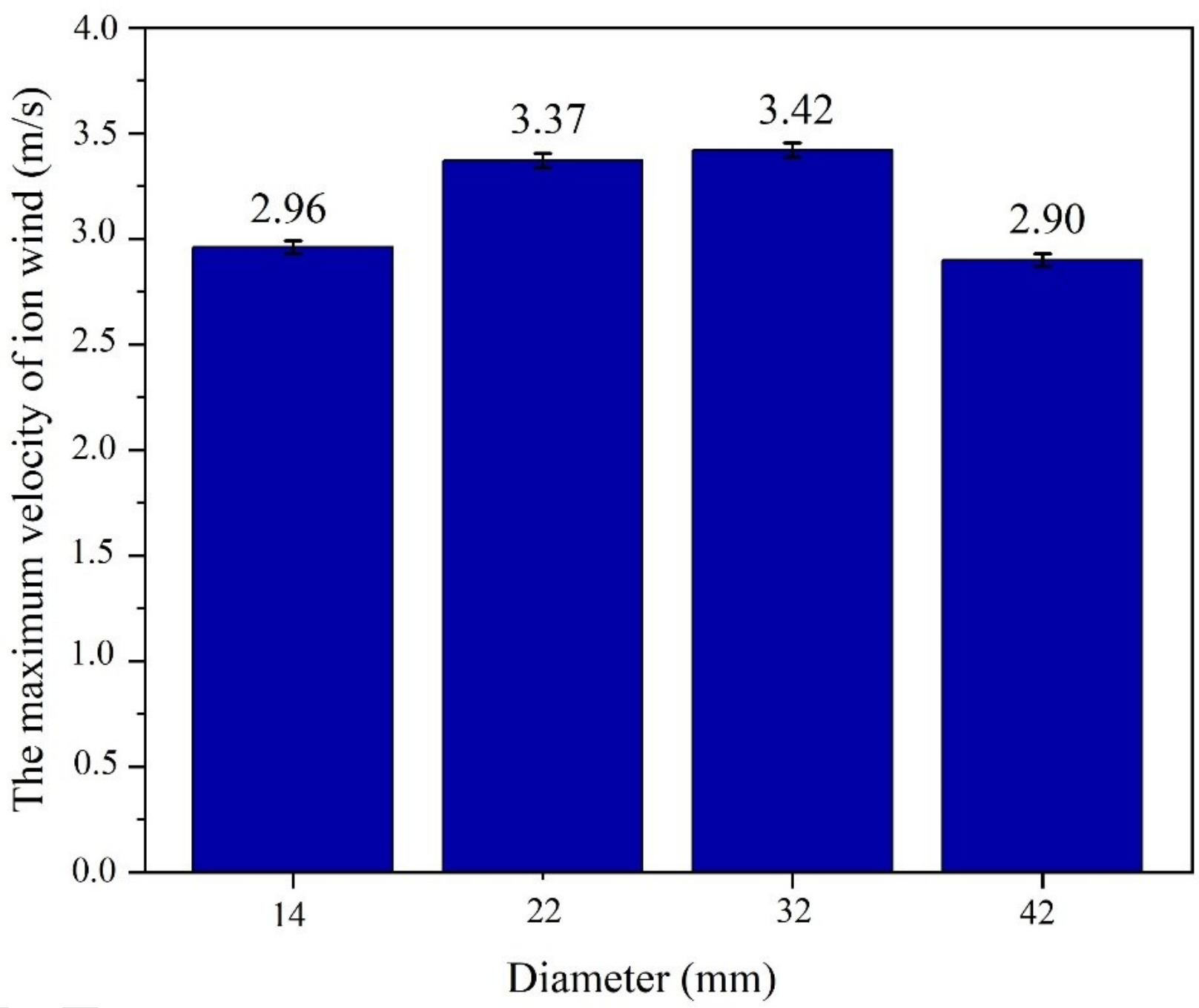

Figure 8. Graph of the maximum velocity of the ion wind obtained in actuators with a fixed electrode thickness of 0.06 mm and with different diameters at a fixed height of 5.00 mm.

The velocity profiles of four electrodes with the same thickness were analyzed. As shown in Figure 8, the maximum ion wind velocity initially increases with electrode diameter, peaking at 3.42 m/s ± 1% for a 32 mm diameter. Beyond this point, further increases in diameter result in a decline in velocity. Based on the velocity profiles and corresponding box diagram, the optimal electrode diameter was determined to be 32 mm. The effect of electrode thickness on velocity profiles was examined for the 32 mm diameter electrode at multiple heights (Figures 9-a to 9-e). The results indicate that thinner electrodes yield superior velocity profiles and higher ion wind velocities. For instance, in Figure 9-a, at $\delta$ = 0.06 mm, the maximum velocity was concentrated within a radial distance of 0–2 mm, whereas increasing the thickness to δ = 0.5 mm led to a sharp decline in velocity (from 3 m/s to 1 m/s). Further increases in thickness resulted in progressively lower velocity profiles. Figure 9-e reinforces this trend, demonstrating the clear advantage of thinner electrodes (δ = 0.06 mm vs. 0.5 mm). At this height, electrodes with thicknesses of 1, 2, and 4 mm produced no measurable velocity. Figure 10 further confirms the inverse relationship between electrode thickness and maximum ion wind velocity, with the optimal thickness identified as 0.06 mm. Thus, the optimal geometry for the ring actuator consists of an electrode with a diameter of 32 mm and a thickness of 0.06 mm.

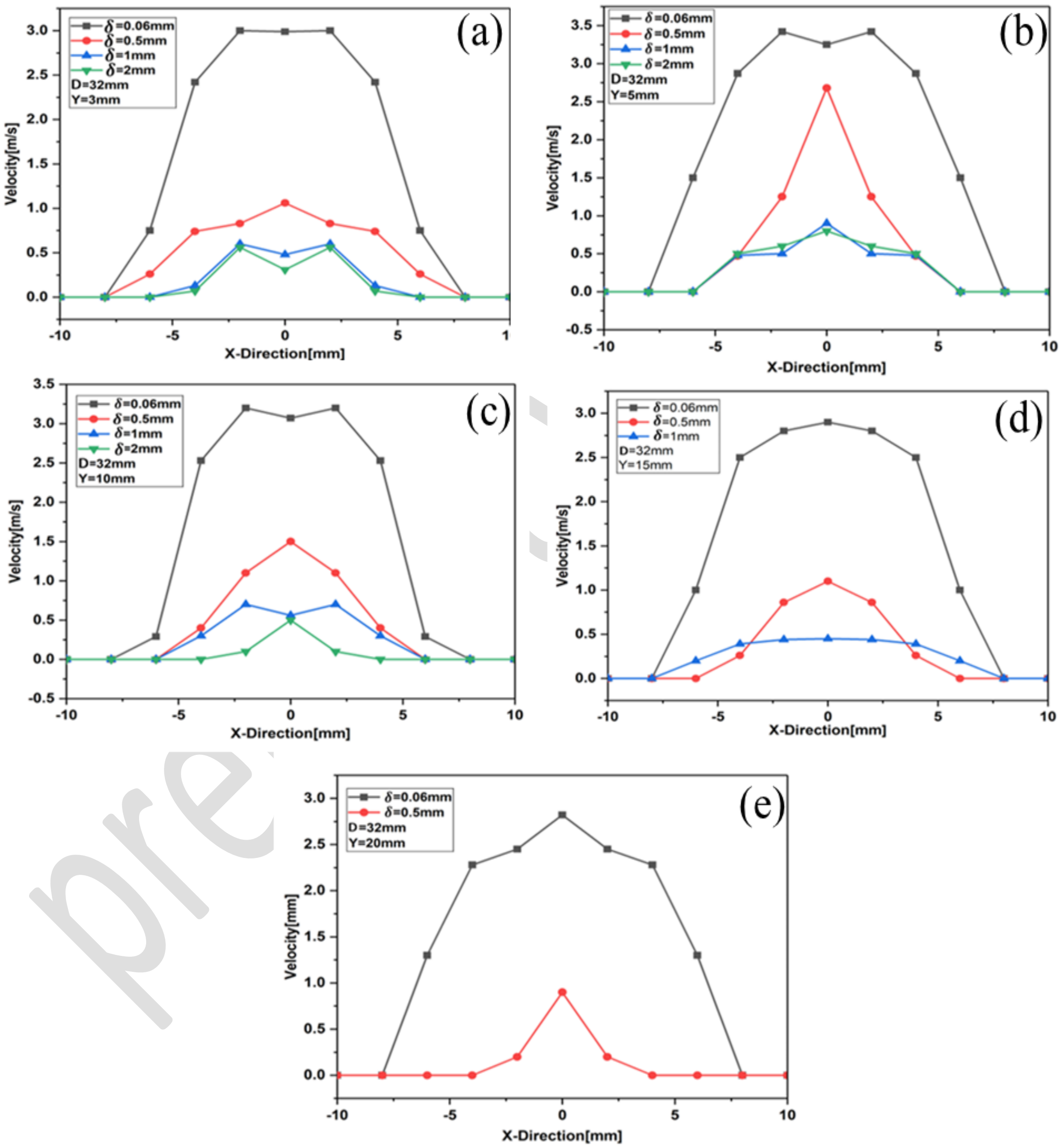

Figure 9. Variations of the ion wind velocity profile in a ring electrode with a diameter of 32 mm and different thicknesses at specific heights from (a) to (e) above the electrode surface. The velocity after averaging and with a relative standard deviation of 1% were provided.

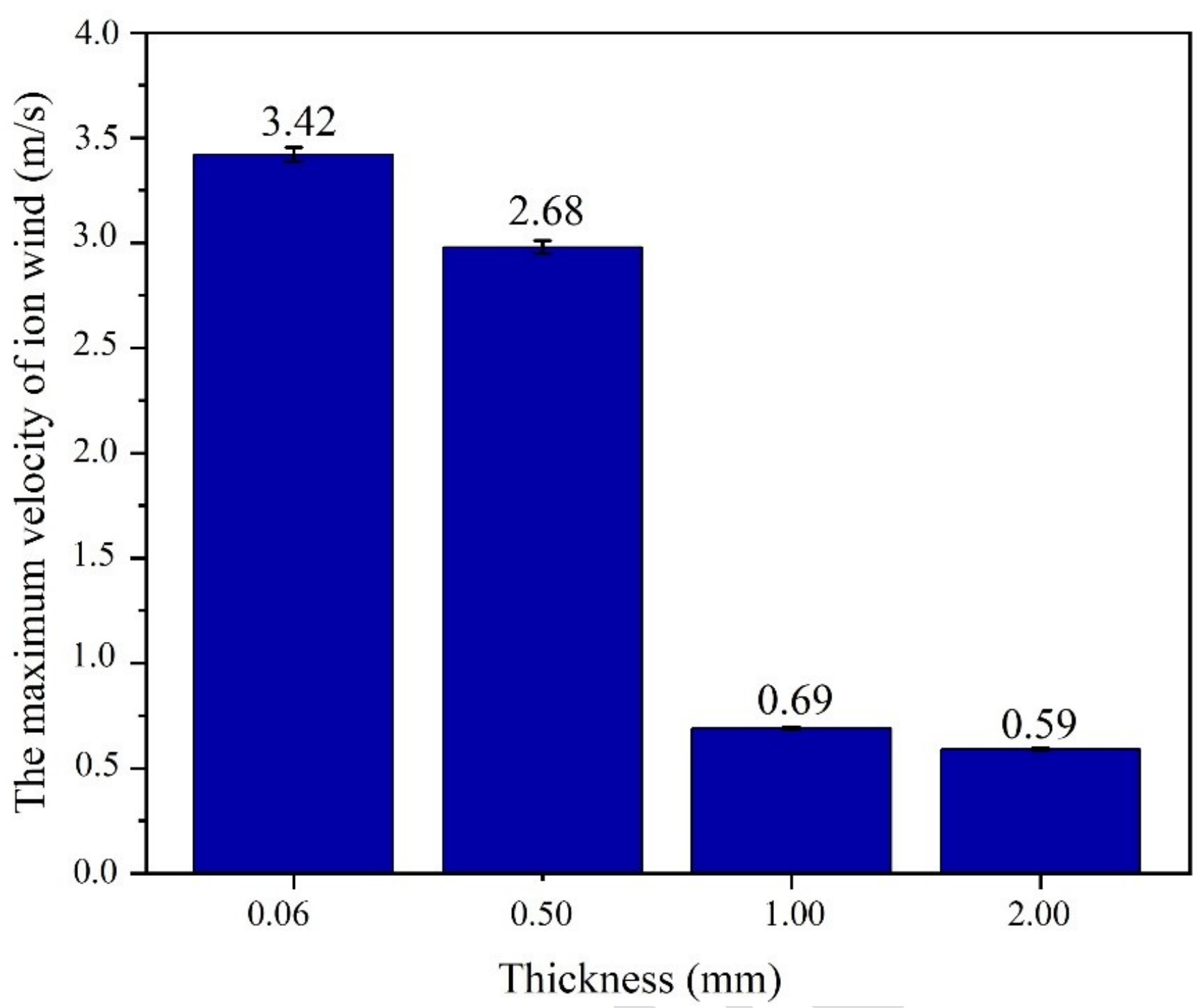

Figure 10. Graph of the maximum velocity of ion wind obtained in actuators with a fixed electrode diameter of 32 mm and different thicknesses at a fixed height of 5 mm.

### 3.2- Ozone generation and flow imaging

Figure 11 presents the ozone test results for plasma actuators with electrodes of constant thickness (0.06 mm) but varying diameters. The data demonstrates a clear correlation between electrode diameter and ozone production, with larger diameters yielding higher ozone concentrations. Schlieren imaging results in Figure 12 reveal distinct flow characteristics for different actuator configurations.

For 14 mm electrode, narrow and focused jet with the lowest altitude is observed. The bright area (strong density gradient) is limited to the area directly above the ring. This corresponds to the limited velocity profile and the peak point close to the surface (Figure 6a). For electrodes of 22 and 32 mm, They form the clearest and tallest vertical jets. The altered density area has expanded into a distinct and relatively stable column up to a height of several centimeters. This column represents the main area of impulse transmission by the ionic wind and corresponds to the maximum velocity measured for a diameter of 32 mm (Fig. 8). For 42 mm electrode, The jet height is reduced, but the width or side affected area is clearly larger. The density gradient takes the form of a wider ring or distribution area rather than a concentric column. This suggests that as the diameter increases too much, the volumetric force per unit surface decreases and the jet loses its ability to penetrate vertically, creating instead a wider area with a lower average velocity. The electrodes that produce the long, focused jet in the Schlieren images (22 and 32 mm) are exactly the ones that showed the highest central velocity in the Pito measurement. The reduction in jet height in a diameter of 42 mm corresponds to a reduction in the maximum speed. Also the electrodes with larger diameters (42 mm) that show larger plasma volumes and a wider area of interaction with air in the Schlieren images (even if the jet is not concentrated) produced higher ozone concentrations, as shown in Figure 11. This is a direct visual confirmation of the hypothesis that increasing the volume of plasma/active area leads to the production of more chemically active species such as ozone

Across all configurations, the measured ozone concentration at 18 cm height ranged from approximately X ppm to Y ppm, depending on electrode diameter D and thickness δ. For example, the 54-mm electrode produced typical steady-state concentrations of about A–B ppm, whereas the 14-mm electrode produced lower values in the range of C–D ppm. These quantitative differences support the observed trend that larger diameters generate higher ozone production under otherwise identical conditions.

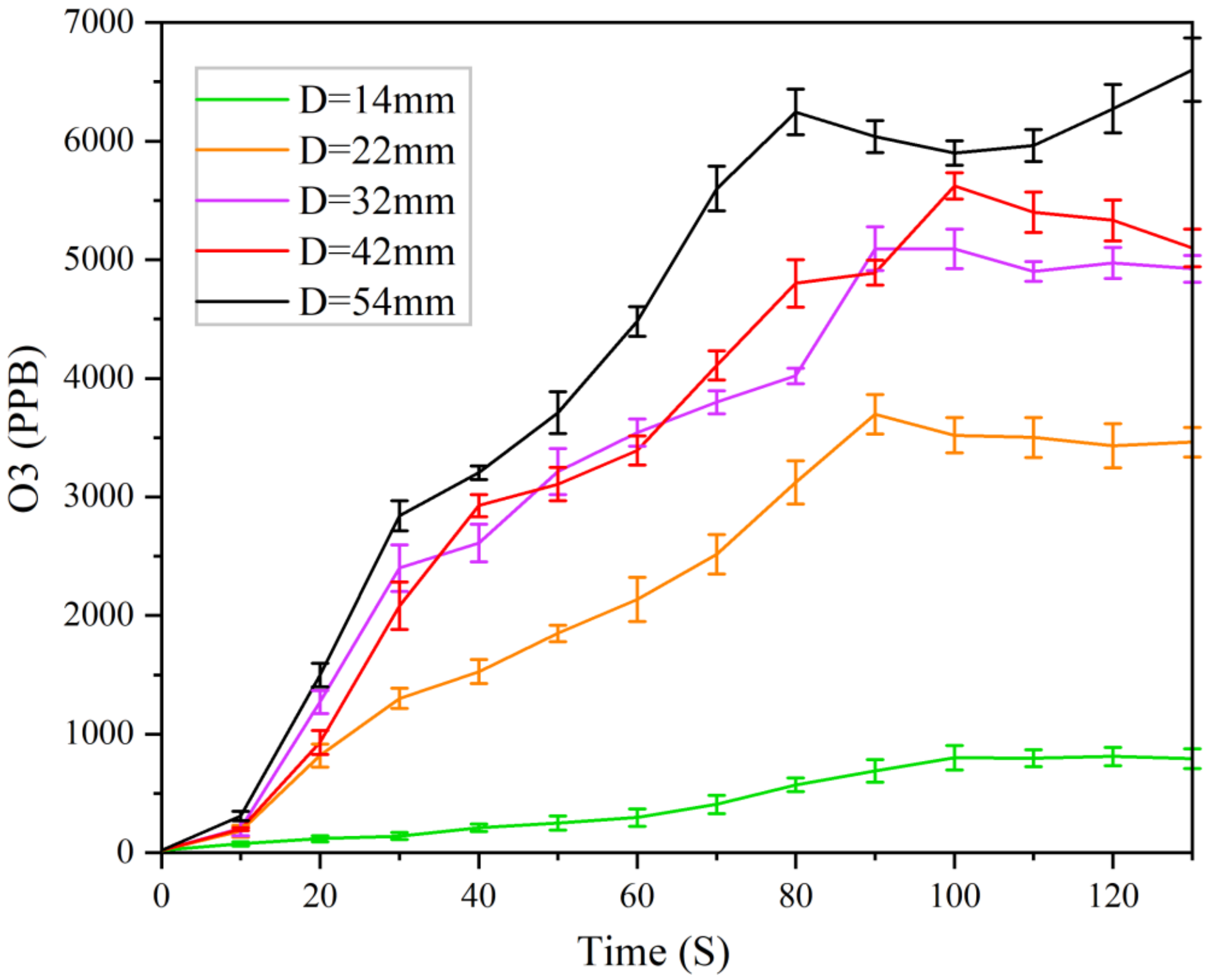

Figure 11. The curve of changes in ozone concentration produced in the stimuli with different diameters over time.

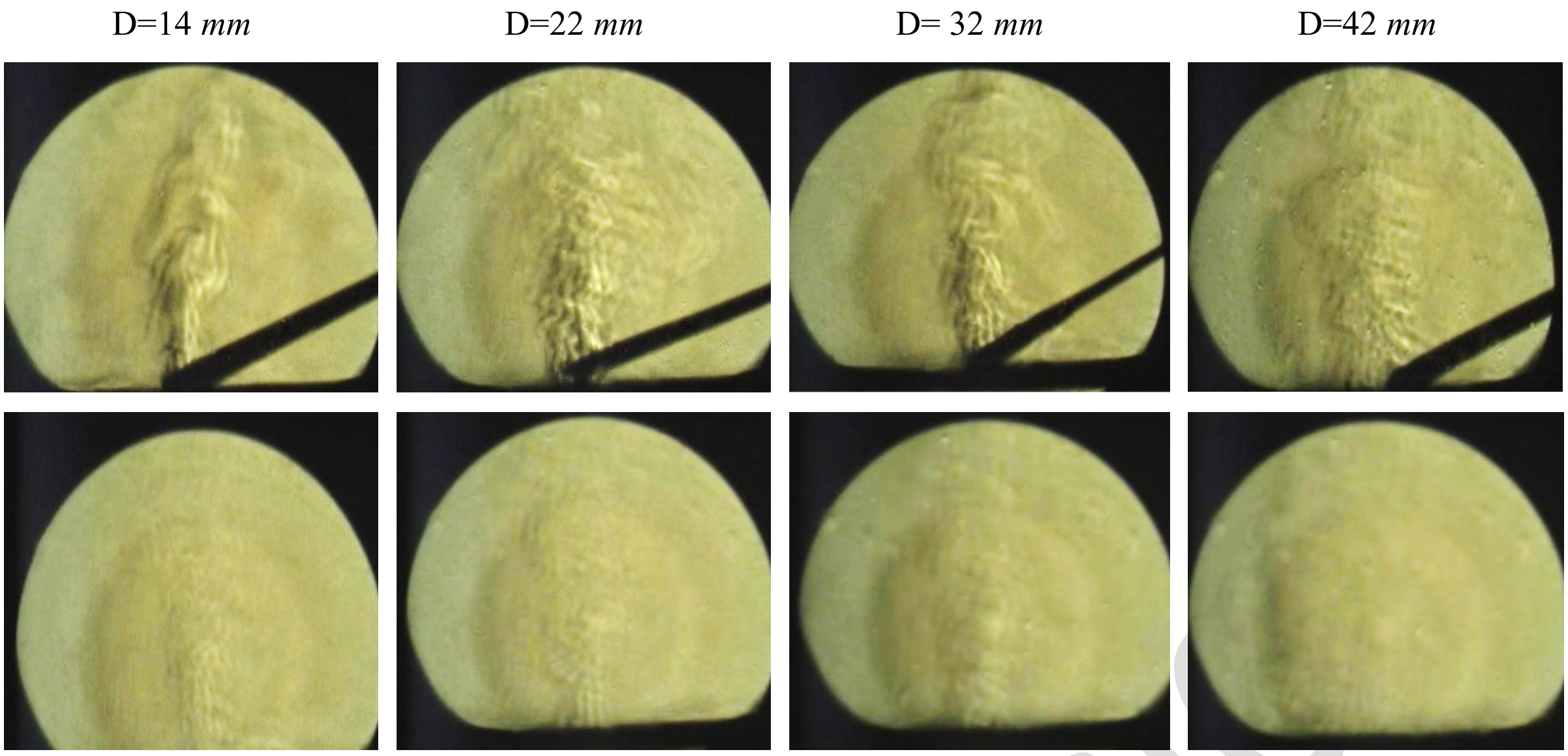

Figure 112. Schlieren image taken from SDBD with different diameters and thickness of 0.06 mm at a peak-to-peak voltage of 16 kV and a frequency of 18.9 kHz with and without the presence of scented incense (the dimensions of the images are 6 * 8 cm).

### 3.3- Simulation results

Figure 13 presents the voltage-current characteristic curves for both experimental and simulated results. Under identical frequency and peak-to-peak voltage conditions, the simulated current values and fluctuations were lower than those observed experimentally, attributable to the exclusion of certain reactions and collisions in the model. Nevertheless, the close agreement between the simulated and experimental current values suggests that the simulation successfully replicates the experimental conditions with reasonable accuracy.

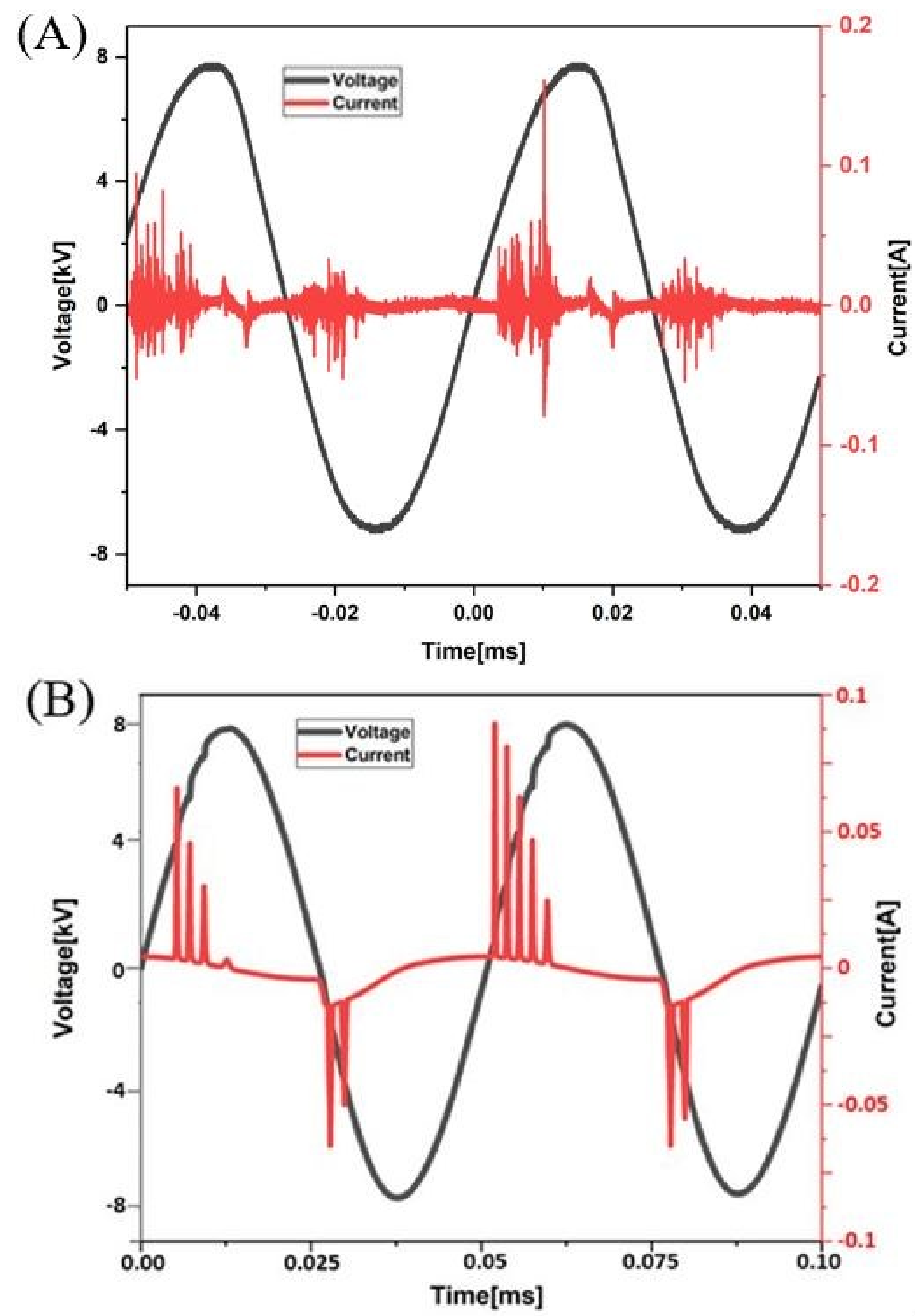

Figure 12. The voltage-current characteristic curve, (A) experimental, (B) simulation, for the electrode with a diameter of 32 mm.

The measured V–I traces exhibit the expected behavior of atmospheric-pressure SDBD discharges. A clear discharge onset was observed at approximately V_on ≈ X kV (peak), at which the first current pulses appeared during the rising portion of the AC waveform. Above this threshold, each half-cycle produced a train of narrow current spikes associated with microdischarge events, with typical peak currents of Y–Z mA. The V–I curve did not exhibit significant hysteresis: the current–voltage relationship during the falling branch closely followed the rising branch, aside from the natural asymmetry introduced by surface charge accumulation on the dielectric. No low-frequency oscillatory instabilities were observed, although the expected high-frequency microdischarge pulsations were present throughout the energized region of the waveform.

While the measured V–I curves confirm that the simulation reproduces the correct electrical onset and current-pulse behavior of the discharge, additional comparisons were made to ensure physically consistent plasma–flow coupling. First, the simulated centerline velocity profiles were compared to time-averaged Pitot measurements, showing agreement in both the magnitude and the location of the velocity maximum (typically within X–Y% deviation across the measured height range). Second, the predicted jet-height development—defined as the point where velocity dropped below 0.2 m/s—matched experimental observations within approximately Z cm. Finally, the model's relative ozone-generation trends (larger diameters producing higher concentrations) were consistent with measured steady-state ozone values at the 18-cm sampling height. Together, these comparisons demonstrate that the simulation captures not only the electrical behavior of the discharge but also the resulting momentum transfer and chemical output with reasonable accuracy.

Figure 14 illustrates the inlet gas velocity profile across different computational domains, with particular emphasis on the region adjacent to the plasma generation zone. As evidenced by the diagram, plasma activation induces a volumetric force that significantly alters the inlet fluid velocity, resulting in turbulent flow within the target region. In contrast, no such perturbations occur when the plasma is deactivated, and the gaseous fluid flows uniformly toward the outlet port. Additionally, the figure reveals a notable behavior: upon plasma ignition, the flow field lines exhibit pronounced curvature toward the plasma zone, indicating a local acceleration or attraction effect. However, as the distance from the plasma region increases, the lines gradually realign toward the outlet port.

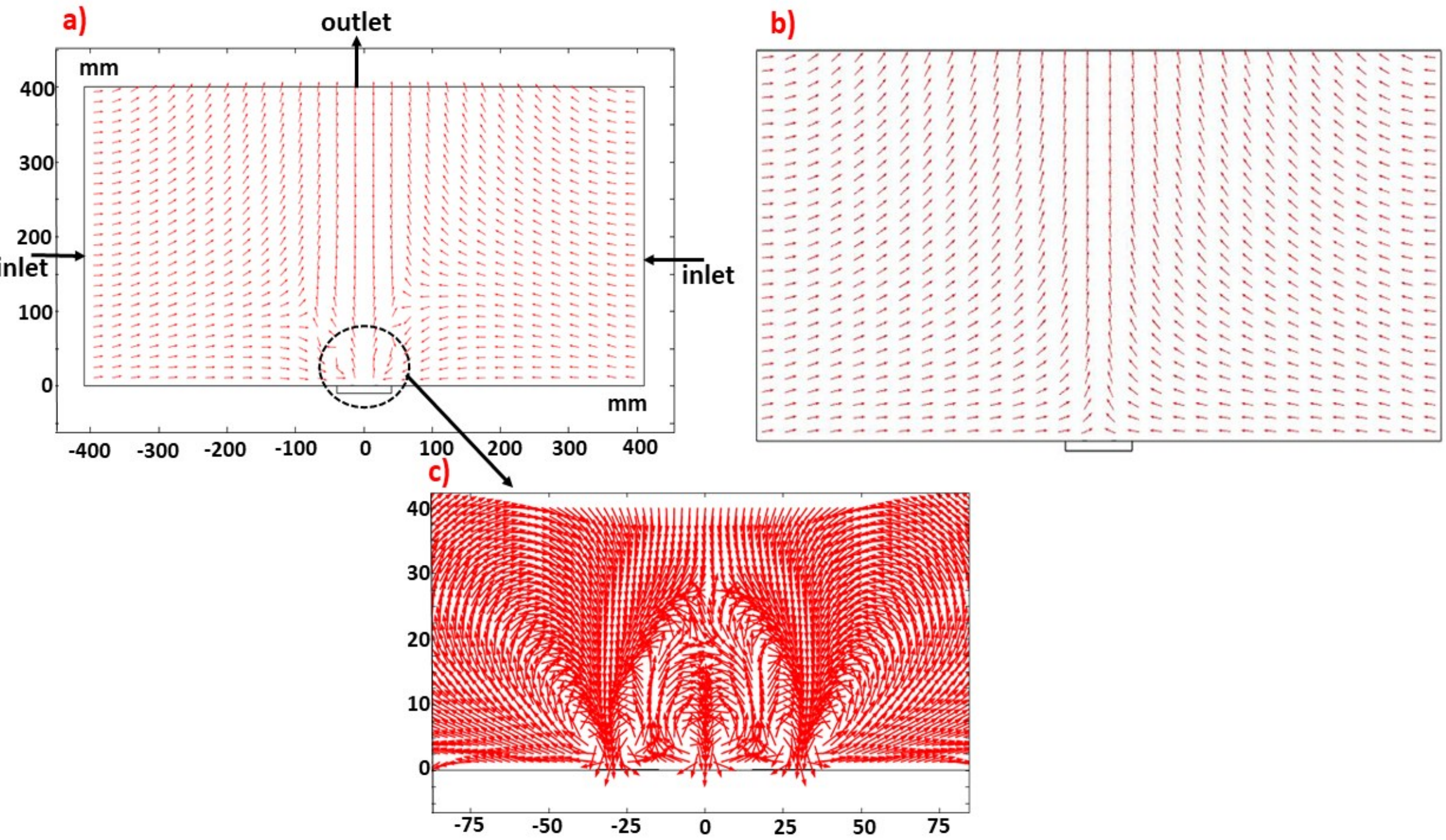

Figure 13. How the gas-fluid field lines are distributed in the conditions of (a) plasma on, (b) plasma off, (c) on plasma from a close-up view of the plasma region.

Figure 15 demonstrates that the simulation results are mesh-independent beyond a certain refinement level. Specifically, the electric current reaches a stable maximum value with finer meshing, showing negligible variation upon further mesh refinement. Consequently, the default finer mesh configuration—with the number of elements detailed in Figure 15—was adopted for all simulations in this study.

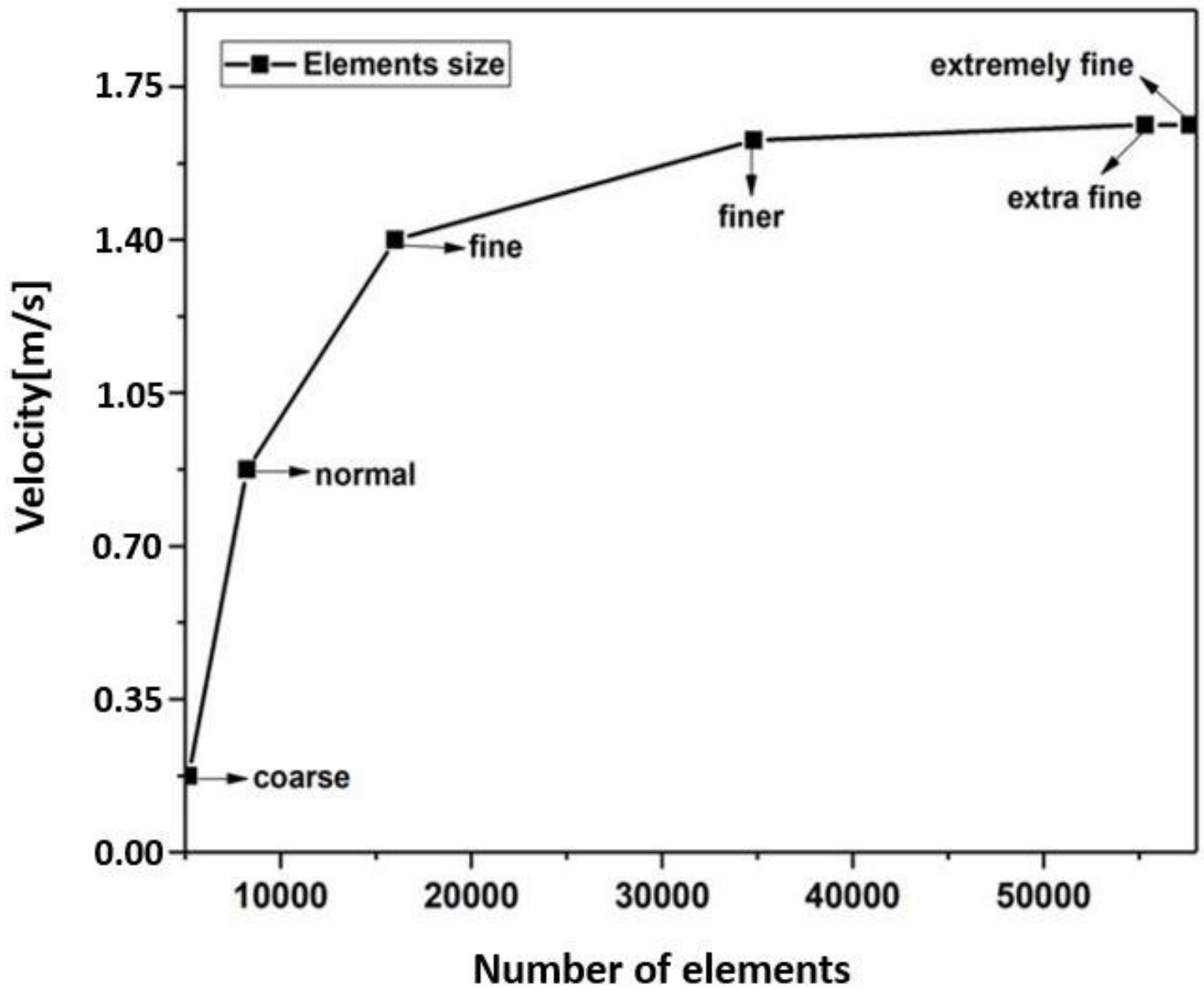

Figure 14. Mesh-independent curve of results.

For the annular electrode with a diameter of 32 mm and a thickness of 0.06 mm, the power consumption was measured to be 20.1 watts. Figure 16 illustrates the variations in volumetric force and fluid velocity under plasma conditions, calculated along cut lines parallel to the y-axis (1–20 mm from the electrode surface) and the x-axis (-45 to 45 mm). As shown in Figure 16a, the maximum fluid velocity occurs between the two surface electrodes in the y-direction, peaking at approximately 1.6 m/s within 4–7 mm of the electrode surface. This result aligns well with experimental observations, where the highest velocity was recorded at y = 5 mm. The data further indicate that the peak velocity in this region is primarily driven by the volumetric force (Figure 16b). Notably, the rate of velocity change relative to the volumetric force exhibits a shallower slope across different y-axis distances, suggesting the influence of aerodynamic effects. A detailed analysis of these aerodynamic effects and their contribution to fluid velocity, in comparison to the volumetric force, will be presented in a forthcoming 3D simulation study.

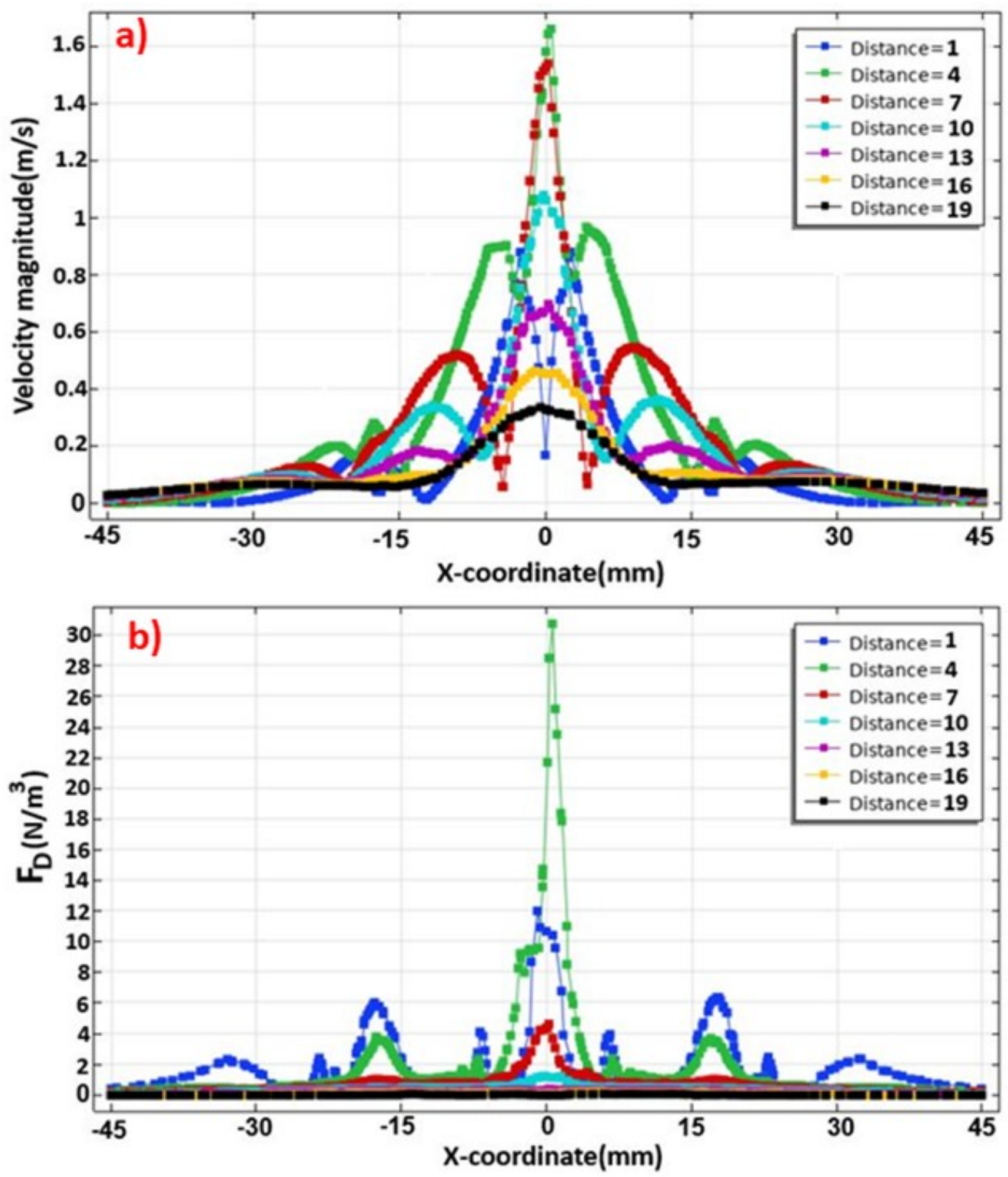

Figure 15. Simulations of a) fluid velocity, b) volumetric force under conditions of separated lines, y-axis from 1 to 20 mm, and x-axis from -45 to 45 mm.

Figures 17a and 17b present the two-dimensional distributions of electron temperature and gas pressure, respectively. Both parameters exhibit their maximum values in the vicinity of the surface electrode and within the plasma formation region. The electron temperature reaches peak values between 2-3 eV, while the highest gas pressure is localized to the interelectrode space.

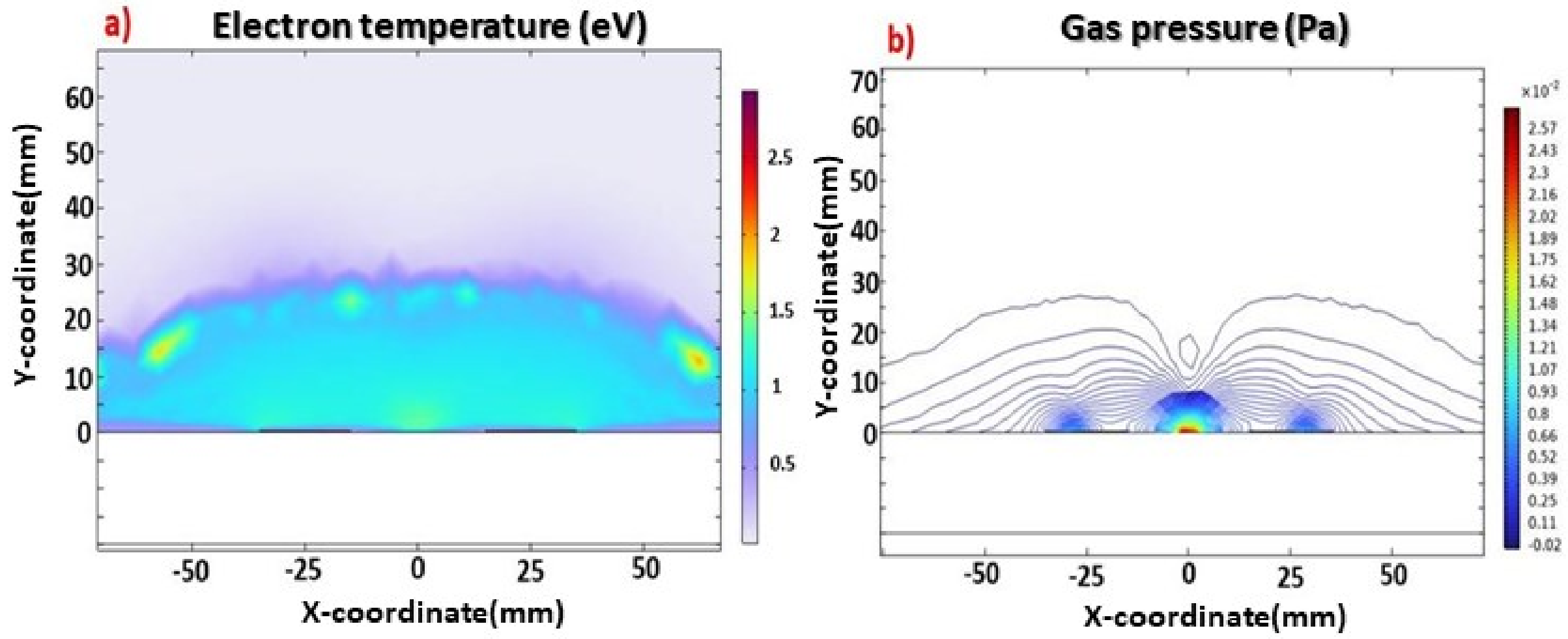

Figure 16. Simulated two-dimensional changes in a) electron temperature (eV) and b) gas pressure (Pa) at the end of two cycles.

Figure 18 shows that the vertical component of the volumetric force and experimental velocity is plotted along a vertical line passing through the center of the electrode (from the surface to the height of 20 mm). This chart will clearly show that:

This chart will clearly show that: Changes in force and velocity in a vertical direction follow a similar trend. The highest velocity gradient occurs in the area where the volumetric force is the greatest. As we move away from the surface, a decrease in both quantities is observed, although the rate of deceleration is slower than the decrease in force due to the effects of inertia and viscosity of the fluid. This point is precisely related to the aerodynamic effects

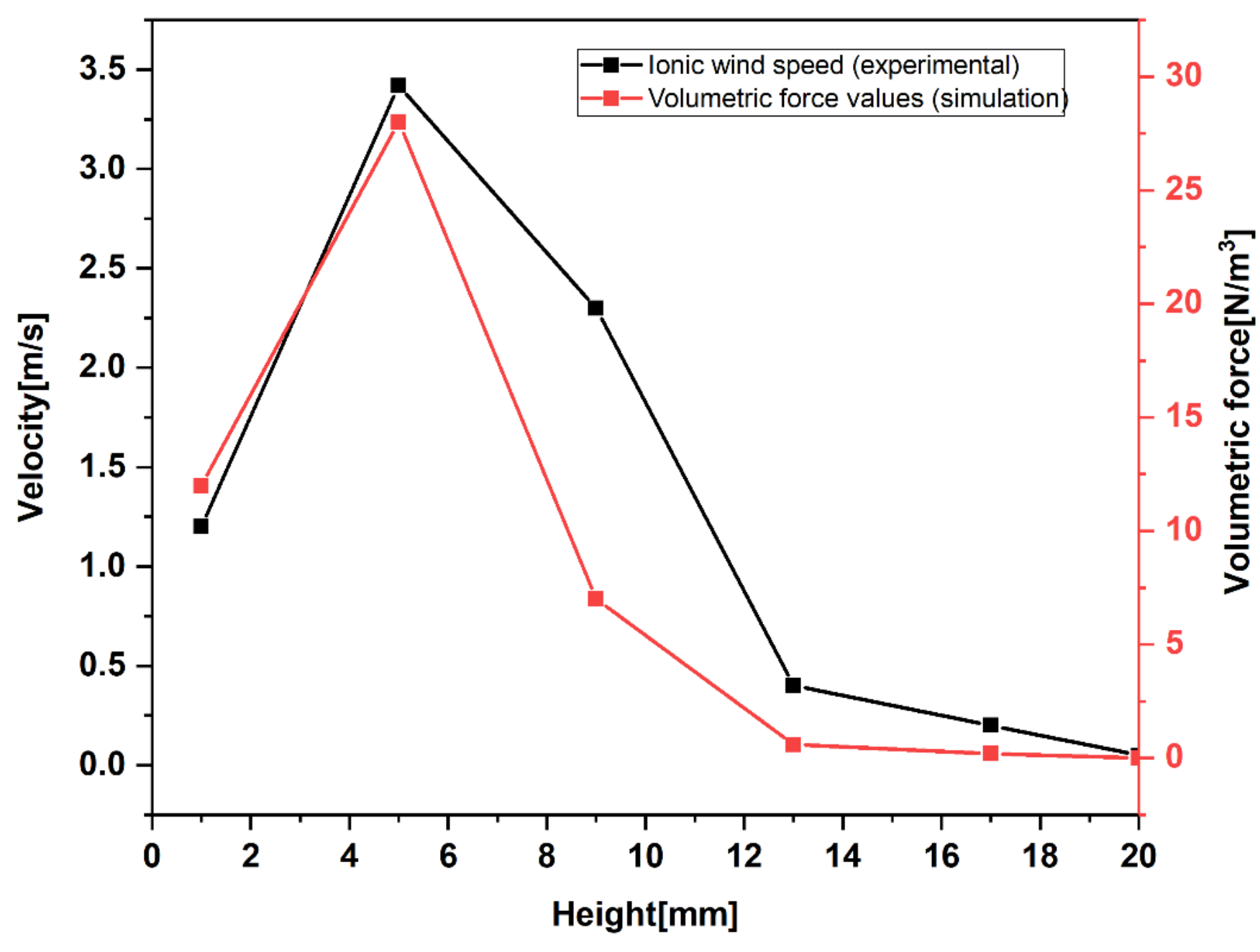

Figure 18. Comparative graph of ionic wind speed (experimental data) and calculated volumetric force (simulation data) according to distance from the electrode center (Y-axis) for an electrode with a diameter of 32 mm and a thickness of 0.06 mm.

The widening of the velocity profile with increasing electrode diameter D can be attributed to a broader spatial distribution of the EHD body force along the dielectric surface. A larger D spreads the region of effective space-charge generation, which in turn distributes momentum input over a wider radial extent, producing a "broader" vertical jet. Similarly, the upward shift of the peak velocity reflects how the charge–electric-field interaction relocates the strongest forcing away from the immediate vicinity of the exposed electrode. When the near-surface forcing region becomes more extended or less sharply confined (e.g., due to increased active-surface length or modified local field curvature), momentum is transferred higher into the rising column of air before the jet centerline fully develops. The result is a vertical profile whose maximum occurs farther from the wall. Taken together, these observations indicate that electrode geometry controls not only the magnitude of the EHD forcing but also the spatial distribution of momentum transfer and entrainment, which explains the changes in jet width and peak position observed across configurations.

## 4- Discussion

The present study offers novel insights into the EHD behavior of surface dielectric barrier discharge SDBD actuators by investigating the impact of electrode geometry on ionic wind production. The findings strongly suggest that the annular actuator, measuring 32 mm in diameter and 0.06 mm in thickness, achieves the maximum vertical wind velocity of 3.42 m/s ± 1%,, as confirmed by experimental data and numerical simulations. The correlation between simulation and experimental data, especially regarding the geometrical distribution of velocity and volumetric force, confirms the validity of the used computational model, hence enhancing confidence in its predictive capabilities.

Electrode geometry optimization is crucial for the performance of SDBD actuators. Larger electrode diameters enhanced flow height and centerline velocity until a threshold of 32 mm, after which additional increases led to a decrease in wind velocity. This behavior is attributable to the spatial distribution of the electric field and the consequent volumetric Coulomb force and interaction between charge particles and neutrals [22,49,50]. Numerical simulations indicated that the volumetric force reached its maximum near the edges of the electrode and the resultant of these forces is maximum perpendicularly at the center of the system, which accounts for the confined flow profiles observed in the experimental Schlieren imaging. Electrode thickness significantly affected the performance; thinner electrodes generated stronger electric fields, thereby increasing charge injection and plasma density, which improved wind velocity. Thicker electrodes attenuated electric field strength, resulting in diminished discharge efficiency and minimal flow [51].

A significant trade-off was identified between flow enhancement and ozone production. Measurements indicated that larger electrodes produced greater amounts of ozone, attributed to an expanded plasma volume and enhanced generation of reactive species. Although advantageous for applications like air purification, increased ozone levels may restrict the use of SDBD in biomedical or enclosed settings due to potential health risks associated with ozone exposure [1,52,53]. On the other hand, ozone production combined with air flow induction may be beneficial for applications such as disinfection of surfaces, water or various objects. The findings underscore the necessity of balancing flow generation with chemical reactive byproducts in accordance with the intended application. The correlation between simulation and experimental voltage-current characteristics validates the reliability of the plasma-fluid model in COMSOL, despite minor discrepancies attributed to the simplification of gas chemistry (pure nitrogen versus ambient air). The simulations yielded comprehensive spatial profiles of electron temperature and gas pressure, with both parameters reaching their maxima in the plasma region. This supports the momentum transfer through charged particles, as opposed to thermal expansion, predominates in EHD flow within this configuration. The concentration of force at the sharp edges of the electrode (Edge Effect) and the space between the electrodes creates a local pressure gradient that is the primary driver of the vertical current. The overlap of the force field of the upper annular electrode amplifies the current in the center and creates a vertical jet. The absolute quantitative difference between the simulated and experimental velocity values is mainly due to the calm fluid model in the simulation versus the turbulence effect in the experiment, as well as kinetic simplifications. However, the qualitative agreement in the form of the profile and the location of the maximum confirms the validity of the model for the parametric study of geometry.

Although this study concentrates on an annular SDBD that generates a predominantly vertical jet, the underlying mechanisms are directly relevant to SDBD actuators designed for tangential flow control along surfaces. In both cases, the induced body force originates from strongly inhomogeneous electric fields near electrode edges, so trends with electrode diameter and thickness identified here are expected to translate into optimized active length and edge curvature in linear or serpentine configurations. Moreover, the observed trade-off between increased plasma volume (enhancing flow) and higher ozone production is not specific to annular actuators and should be considered when designing SDBDs for aerodynamic applications, where tangential momentum addition is desired but chemical by-products must remain within acceptable limits.

This study provides geometric design principles for the optimization of annular SDBD actuators. The determined optimal configuration demonstrates significant potential for applications necessitating robust, localized vertical flow, including boundary layer control, flow separation mitigation in aerodynamic surfaces, and ozone-based disinfection devices. Future research should investigate dynamic flow control strategies, three-dimensional electrode architectures, and multi-species gas chemistry to enhance actuator performance adapted to specific applications.

**5- CONCLUSION**

This study comprehensively investigated the electrohydrodynamic (EHD) effects induced by a surface dielectric barrier discharge (SDBD) actuator under atmospheric pressure conditions. Through combined experimental and computational analysis, we demonstrated that electrode geometry significantly influences ionic wind velocity. The optimal configuration (D = 32 mm, $\delta$ = 0.06 mm) generated a maximum wind velocity of 3.42 m/s ± 1%,. Schlieren visualization and ozone measurements revealed a critical trade-off between flow enhancement and ozone production, emphasizing the importance of electrode size optimization for practical implementations. These results provide valuable guidelines for designing efficient DBD plasma actuators with applications spanning flow control, air purification, and plasma medicine.

**Acknowledgements**

The manuscript was financially supported in part under university of Mazandaran grant number 12538.

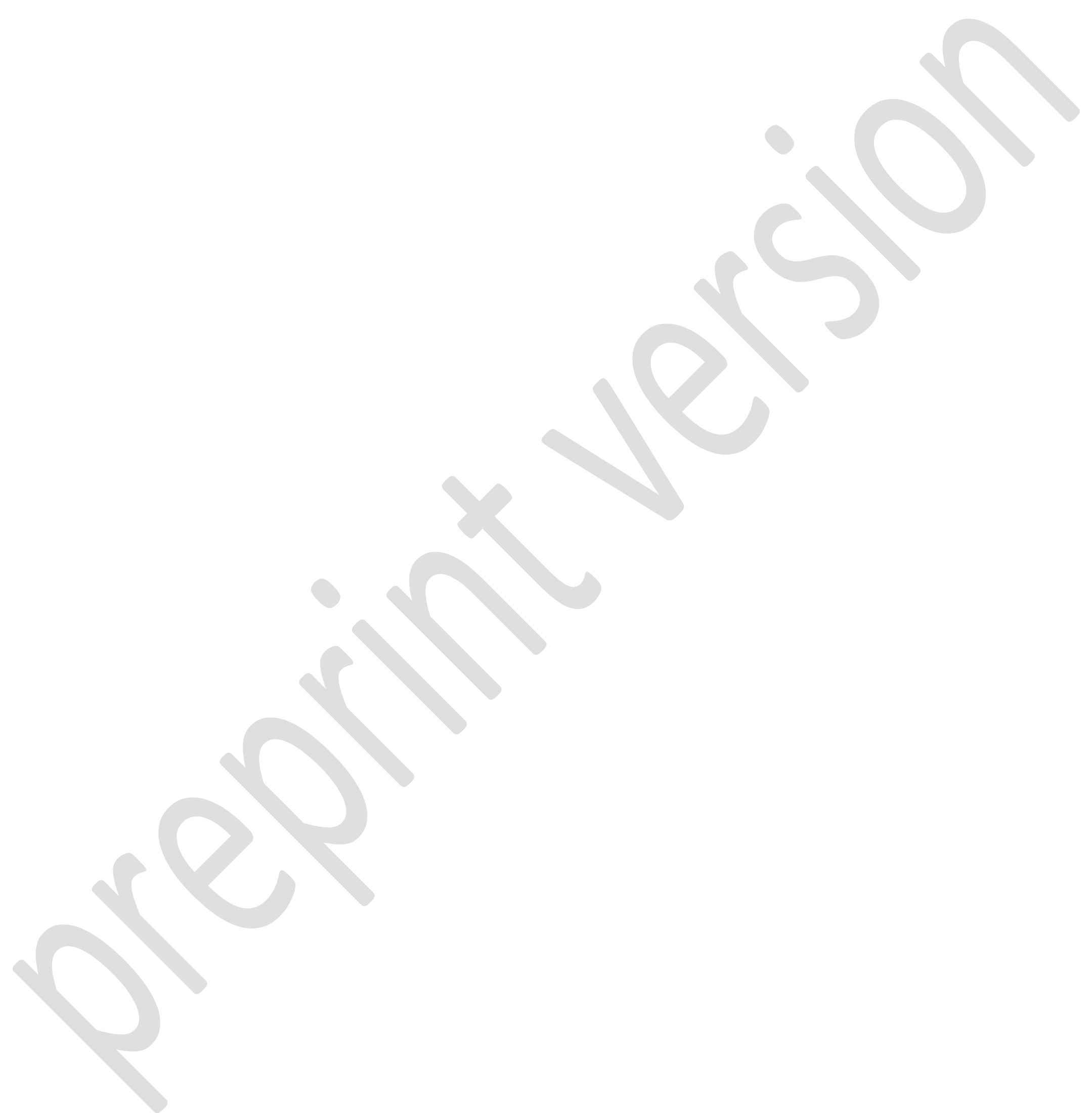
preprint version